\documentclass[10pt,twoside,twocolumn,english,aps,nofootinbib,preprintnumbers,superscriptaddress]{revtex4}
\usepackage[latin9]{inputenc}
\usepackage{color}
\usepackage{babel}
\usepackage{mathrsfs}
\usepackage{amsmath}
\usepackage{amssymb}
\usepackage{esint}
\usepackage[unicode=true,pdfusetitle,
 bookmarks=true,bookmarksnumbered=false,bookmarksopen=false,
 breaklinks=true,pdfborder={0 0 0},backref=false,colorlinks=true]
 {hyperref}

\makeatletter
\@ifundefined{textcolor}{}
{%
 \definecolor{BLACK}{gray}{0}
 \definecolor{WHITE}{gray}{1}
 \definecolor{RED}{rgb}{1,0,0}
 \definecolor{GREEN}{rgb}{0,1,0}
 \definecolor{BLUE}{rgb}{0,0,1}
 \definecolor{CYAN}{cmyk}{1,0,0,0}
 \definecolor{MAGENTA}{cmyk}{0,1,0,0}
 \definecolor{YELLOW}{cmyk}{0,0,1,0}
}

\usepackage{hyperref}
\hypersetup{
    colorlinks,%
    citecolor=blue,%
    filecolor=blue,%
    linkcolor=blue,%
    urlcolor=blue
}

\makeatother

\begin{document}

\preprint{CERN-PH-TH/2012-205}
\preprint{\emph{published in} EPL, 101 (2013) 34001}

\title{Suppressing Quantum Fluctuations in Classicalization}

\author{Alexander Vikman}

\affiliation{CERN, Theory Division, CH-1211 Geneva 23, Switzerland}

\email{alexander.vikman@cern.ch}

\selectlanguage{english}%

\date{\today}
\begin{abstract}
We study vacuum quantum fluctuations of simple Nambu-Goldstone bosons
- derivatively coupled single scalar-field theories possessing shift-symmetry
in field space. We argue that quantum fluctuations of the interacting
field can be drastically suppressed with respect to the free-field
case. Moreover, the power-spectrum of these fluctuations can soften
to become \textit{red} for sufficiently small scales. In \textit{quasiclassical}
approximation, we demonstrate that this suppression can only occur
for those theories that admit such \textit{classical static} backgrounds
around which small perturbations propagate faster than light. Thus,
a \textit{quasiclassical softening} of quantum fluctuations is only
possible for theories which \textit{classicalize} instead of having
a usual Lorentz invariant and local Wilsonian UV- completion. We illustrate
our analysis by estimating the quantum fluctuations for the DBI-like
theories.
\end{abstract}
\maketitle

\section{Introduction}

Quantum Mechanics (QM) of a finite number of degrees of freedom is
a paradigm which should a priory be applicable to arbitrary Hamiltonian
systems including strongly coupled models. In particular, QM describes
systems with noncanonical and even highly nonlinear kinetic terms
as well as it works for those with the highly nonlinear potentials.
However, currently, in the case of QM for a continuum number of degrees
of freedom (i.e. for quantum field theory (QFT)) all models well defined
at all scales (renormalizable) are just quadratic in canonical momentum.
For QFT a noncanonical structure of the kinetic term usually implies
the existence of a strongly coupled regime where the usual perturbative
renormalization procedure breaks down and where one has to use non-perturbative
methods, e.g. lattice computation. However, lattices of different
size correspond to the approximation of QFT by finite dimensional
QM systems with different number of degrees of freedom and the convergence
of this procedure is not obvious. 

It is a common belief that Nature is such that it can be described
in terms of fields which are different in the weak and strong coupling
regimes. The best examples for this behavior are given within the
Standard Model by QCD and Higgs field in the Electroweak sector. One
tends to think that in UV the number of degrees of freedom - types
of particles - should increase, see e.g. most recent works \cite{Komargodski:2011vj,Luty:2012ww}.
New particles should be integrated in to provide a UV complete theory
- this is the basis of the Wilsonian UV-completion. This is the case
for String Theory where in the UV there is a continuum of fields.
In this regard natural questions arise: is continuum QM - QFT -infinitely
more restrictive than QM for finite number of degrees of freedom?
Can noncanonical or perturbatively non-renormalizable QFTs make sense
in a continuum limit i.e. for \textit{all scales}? If it is the case,
can one understand the mechanism used by Nature in these theories
in terms of some weakly coupled fields / particles? These questions
could be crucial for understanding of Quantum Gravity which has a
noncanonical Hamiltonian with the strong-coupling scale given by the
Planck mass. 

Recently it was proposed \cite{Dvali:2010bf,Dvali:2010jz,Dvali:2010ns,Dvali:2010ue}
that gravity may be self-UV-completed via \textit{classicalization}
- the softening of quantum fluctuations i.e. loops in Feynman diagrams
at the transplanckian transferred momenta, because of the formation
of intermediate Black Holes\footnote{For an earlier similar discussion of the quasiclassical high-energy scattering in gravity, see e.g. \cite{Banks:1999gd}.}. \textit{Classicalization} is caused by
the high level of nonlinearity and corresponding self-sourcing. Moreover, it was suggested that this mechanism may also work for
quite generic class of the Nambu-Goldstone bosons \cite{Dvali:2011th,Dvali:2010ns}\footnote{For criticism of this interesting proposal, see e.g. Refs. \cite{Akhoury:2011en,Kovner:2012yi}} and even more general scalar fields including nonlinear sigma models
\cite{Percacci:2012mx}, where the role of Black Holes is played by
\textit{classicalons} - extended and long lived classical configurations.
However, this phenomenon was argued \cite{Dvali:2011nj,Dvali:2012zc}
to occur for those models for which the usual local and Lorentz-invariant
Wilsonian UV-completion is absent, because they allow for a superluminal
propagation of perturbations around nontrivial backgrounds \cite{Adams:2006sv}
\footnote{Here it is interesting to note that already in 1952 Heisenberg argued
that derivative self-interaction is important for the correct description
of high multiplicity in scattering in strongly interacting theories.
See \cite{Heisenberg:1952zz} where the usual subluminal DBI is used
to analyse this problem. %
}.

In this paper we analyse the selfconsistency of \textit{classicalization}
for general noncanonical Nambu-Goldstone bosons in $d+1$ spacetime
dimensions. We did not restrict our attention to $d=3$, because in
lower spacial dimensions there are much higher chances to find theories
for which our estimations can be checked by numerical simulations
or by other non-perturbative methods. We use the Heisenberg uncertainty
relation for canonically conjugated field and momentum operators averaged
on a given scale to estimate vacuum quantum fluctuations of the field
on this scale. In our analysis we follow the ideas of \textit{classicalization}
and work in the \textit{quasiclassical} approximation - neglecting
higher-order correlators. On top of that we use analytic continuation
which is a standard tool for interacting QFT with the energy functional
bounded from below. We show that vacuum quantum fluctuations, $\delta\phi$, can be
suppressed with respect to the free-field case only if the theory
admits \textit{static classical} backgrounds which are either i) absolutely
unstable (have gradient instability) or ii) such that perturbations
around them propagate faster than light. Option i) is definitely even
more worrisome than ii) and should be excluded. Moreover, if the suppression
is such that the spectrum of perturbations is \textit{red} in the
UV, then the speed of perturbations $v^{2}>d\geq1$. We illustrate
our finding by analysis of quantum fluctuations for Dirac-Born-Infeld
(DBI) like theories. 

\section{Heisenberg uncertainty relation for averaged fields }

We are interested in the characteristic quantum fluctuation $\delta\phi\left(\ell\right)$
on scale $\ell$. In four spacetime dimensions, and weakly coupled
canonical field theories the standard result is 
$\delta\phi\left(\ell\right)\simeq\sqrt{\hbar}/\ell$,
for scales much shorter than the possible Compton wavelengths in the theory.
Following the \textit{classicalization} idea, we assume that even
for the lengths smaller than the strong coupling scale $M_{\text{s}}^{-1}$,
the field operator, $\hat{\phi}\left(x\right)$, and the canonically
conjugated momentum $\hat{p}\left(x\right)$, can still be defined%
\footnote{To simplify notation, we will mostly omit hats indicating the operators
and write the action quantum $\hbar$ only in some formulas to stress
their quantum origin, otherwise the system of units is such that $c=\hbar=1$. %
}. Also we assume that one can define states $\left|\Psi\right\rangle $$ $
and in particular a vacuum state, $\left|0\right\rangle $, applicable
for \textit{all }scales for the \textit{interacting} theory. We define
the operators averaged%
\footnote{Note that an averaging in time is not needed, because one can equally
well consider the system in the Schr\"odinger picture where all operators
are time-independent. Moreover, for any energy eigenstate $\left|E\right\rangle $
the expectation values for all observables are time-independent. %
} on scale $\ell$ as 
\begin{equation}
\hat{\phi}_{\ell}\left(\mathbf{x},t\right)=\int\mbox{d}^{d}\mathbf{x}'\, W_{\ell}^{\phi}\left(\mathbf{x}-\mathbf{x}'\right)\hat{\phi}\left(\mathbf{x}',t\right)\,,\label{eq:Averaging}
\end{equation}
where $W_{\ell}^{\phi}\left(\mathbf{x}-\mathbf{x}'\right)$ is a window
function centered at $\mathbf{x}$. It is convenient and common to
assume that a device measuring the scalar field can operate with adjustable
(but always finite!) resolution (averaging on different scales $\ell$)
in such a way that a family of window-functions characterising this
device scales as 
\begin{equation}
W_{\ell}^{\phi}\left(\mathbf{x}\right)=\ell^{-d}\cdot w^{\phi}\left(\frac{\mathbf{x}}{\ell}\right)\,,\label{scaling property}
\end{equation}
see e.g. \cite{Mukhanov:2007zz}, where the shape function $w^{\phi}$ is positive with $\int\mbox{d}^{d}\mathbf{r}\, w^{\phi}\left(\mathbf{r}\right)=1$.
For simplicity one can assume that $w^{\phi}\left(\mathbf{r}\right)=w^{\phi}\left(-\mathbf{r}\right)$.
Every single measurement of the scalar field by a device located at
position $\mathbf{x}$ gives an eigenvalue of the operator (\ref{eq:Averaging}).
In the state $\left|\Psi\right\rangle $ the field fluctuation on
length $\ell$ is defined as 
\begin{equation}
\delta\phi_{\ell}^{2}\equiv\left\langle \Psi\right|\hat{\phi}_{\ell}^{2}\left|\Psi\right\rangle -\left\langle \Psi\right|\hat{\phi}_{\ell}\left|\Psi\right\rangle ^{2}\,.
\end{equation}
A device measuring the canonical momentum $\hat{p}\left(x\right)$
will operate with a different window function family $W_{\ell}^{p}\left(\mathbf{x}\right)$
possessing the same natural scaling property (\ref{scaling property}) with some shape function $w^{p}\left(\mathbf{r}\right)$.
The canonical commutation relation is 
\begin{equation}
\left[\hat{\phi}\left(t,\mathbf{x}\right),\hat{p}\left(t,\mathbf{y}\right)\right]=i\hbar\,\delta\left(\mathbf{x}-\mathbf{y}\right)\,,
\end{equation}
which for the operators averaged on the \emph{same scale} $\ell$
gives 
\begin{equation}
\left[\hat{\phi}_{\ell}\left(t,\mathbf{x}\right),\hat{p}_{\ell}\left(t,\mathbf{y}\right)\right]=i\hbar\cdot\ell^{-d}\cdot\mathscr{D}\left(\frac{\mathbf{x}-\mathbf{y}}{\ell}\right)\,,
\end{equation}
where the function $\mathscr{D}\left(\mathbf{r}\right)$ is given
by a convolution of the shape functions 
\begin{equation}
\mathscr{D}\left(\mathbf{r}\right)=\int\mbox{d}^{d}\mathbf{r}'\, w^{\phi}\left(\mathbf{r}-\mathbf{r}'\right)w^{p}\left(\mathbf{r}'\right)\,.
\end{equation}
Now we can use the uncertainty relation in the form derived by Robertson
\cite{PhysRev.34.163} to obtain 
\begin{equation}
\delta\phi_{\ell}\left(t,\mathbf{x}\right)\cdot\delta p_{\ell}\left(t,\mathbf{y}\right)\geq\frac{\hbar}{2}\cdot\mathscr{D}\left(\frac{\mathbf{x}-\mathbf{y}}{\ell}\right)\cdot\ell^{-d}\,.
\end{equation}
To measure the same set of degrees of freedom one would like to measure
the field and conjugated momentum possibly at the same point in space
with the precision \emph{better} than $\ell$ so that at the same
point we have
\begin{equation}
\delta\phi_{\ell}\left(t,\mathbf{x}\right)\cdot\delta p_{\ell}\left(t,\mathbf{x}\right)\geq\frac{\hbar}{2}\cdot\mathscr{D}_{0}\cdot\ell^{-d}\,,\label{HeisenbergUR-1}
\end{equation}
where the number %
\footnote{For example, $\mathscr{D}_{0}=\left(2\sqrt{\pi}\right)^{-d}$ if both
window functions are just Gaussian 
\begin{equation}
W_{\ell}\left(\mathbf{x}\right)=\frac{1}{\left(2\pi\right)^{d/2}\ell^{d}}\exp\left(-\frac{\left|\mathbf{x}\right|^{2}}{2\ell^{2}}\right)\,.
\end{equation}
} $\mathscr{D}_{0}\equiv\mathscr{D}\left(0\right)=\mathcal{O}\left(1\right)>0$,
depends on the exact properties of the window functions (corresponding
to the measuring devices) but does not depend on the scale of averaging
or on the quantum state $\left|\Psi\right\rangle $. An estimation
for the product of fluctuations for averaged fields can be found in
earlier works, e.g. in \cite{Garriga:2001ch}. 

The uncertainty relation (\ref{HeisenbergUR-1}) unambiguously shows
that \emph{the shorter is the scale of averaging the more quantum is the field
theory}. However, if the theory is non-canonical then the field can
have a ``fake quasiclassical'' property that $\delta\phi_{\ell}\times\delta\dot{\phi}_{\ell}\rightarrow0$.
Further, one can remark that in standard
canonical theories for a multi-particle state $\left|N\right\rangle $
which is an eigenstate of energy the product of fluctuations is larger
than that for the vacuum $\left|0\right\rangle $, while $\left\langle N\right|\hat{\phi}_{\ell}\left|N\right\rangle =\left\langle N\right|\hat{p}_{\ell}\left|N\right\rangle =0$,
see e.g. \cite{Mukhanov:2007zz}. Thus large number of particles is
not a sufficient condition for a truly quasiclassical behaviour. 

In the next section we use the uncertainty relation for a scalar field
in the Lorentz-invariant vacuum state $\left|0\right\rangle $ to
estimate quantum fluctuations $\delta\phi_{\ell}\left(\mathbf{x}\right)$.
In the Lorentz-invariant vacuum the fluctuations are the same at all
points of the space, so that $\delta\phi_{\ell}\left(\mathbf{x}\right)=\delta\phi\left(\ell\right)$.
An attractive interaction reduces $\delta\phi\left(\ell\right)$ with
respect to the free-field case while a repulsive interaction increases
$\delta\phi\left(\ell\right)$. It is well known (see e.g. \cite{Mukhanov:2007zz})
that $\delta\phi^{2}\left(\ell\right)$ corresponds to the normalised
power spectrum for the two-point correlation function, $\left\langle 0\right|\hat{\phi}\left(t,\mathbf{x}\right)\hat{\phi}\left(t,\mathbf{y}\right)\left|0\right\rangle $,
where $\ell=\left|\mathbf{x}-\mathbf{y}\right|$$ $.

\section{Vacuum Quantum Fluctuation for simple Nambu-Goldstone bosons}

Here we consider a general theory with shift symmetry, $\phi\rightarrow\phi+c$,
in $d+1$ spacetime dimensions with the action 
\begin{equation}
S=\int\mbox{d}^{1+d}x\,\mathscr{L}\left(X\right)\,, \ \ \text{where\footnote{We use the signature $\left(+---\right)$}} \ \ \ X=\frac{1}{2}\left(\partial\phi\right)^{2} \ ,\label{ActionGeneral}
\end{equation}
denotes the standard kinetic term. We assume that this QFT has a strong
coupling scale $M_{\text{s}}^{-1}$ which is supposed to be \emph{the
only characteristic length} in the theory. $ $The canonical momentum
is 
\begin{equation}
p=\frac{\partial\mathscr{L}}{\partial\dot{\phi}}=\mathscr{L}_{X}\dot{\phi}\,,
\end{equation}
where $\mathscr{L}_{X}\equiv d\mathscr{L}/dX$. The fluctuation of
the momentum can be estimated as 
\begin{align}
& \delta p\left(\ell\right)=\sqrt{\left\langle 0\right| \left(\mathscr{L}_{X}\dot{\phi}\right)_{\ell}^{2}\left| 0\right\rangle}=\delta p_{\text{g}}\left(\delta\phi_{\ell}/\ell\right)+\label{eq:assumption}\\
& +\text{higher order correlators}\,.\nonumber 
\end{align}
Following the idea of \textit{classicalization} we assume that the
theory is indeed \textit{``weakly} \emph{coupled}'' in deep UV so
that one can neglect higher order correlators. In particular, this
is true for an approximately Gaussian vacuum state. Of course, the
original particles are strongly coupled and form extended multiparticle
objects - \emph{classicalons}. The later can be considered as \emph{weakly
coupled} \emph{quasiparticles} in UV. Moreover, the shorter is the
scale the more classical is the behavior of $\phi$. Using the quasiclassical
approximation $\left\langle \!\right.f(\mathcal{\hat{O}})\left.\!\right\rangle \simeq f(\left\langle \!\right.\mathcal{\hat{O}}\left.\!\right\rangle )$
one can argue that function $\delta p_{\text{g}}$ can be estimated
as 
\begin{equation}
\delta p_{\text{g}}\left(\delta\phi_{\ell}/\ell\right)\simeq\mathscr{L}_{X}\left(X_{\ell}\right)\delta\phi_{\ell}/\ell\,,
\end{equation}
where the characteristic $X_{\ell}$ is 
\begin{equation}
X_{\ell}=-\left(\delta\phi_{\ell}/\ell\right)^{2}\,.\label{X}
\end{equation}
In this estimation we could write in front a scale independent (or
slowly $\ell$-dependent) number $\gamma=\mathcal{O}\left(1\right)>0$,
however it can always be absorbed into the normalisation of $\delta\phi$
and $\mathscr{D}_{0}$. Here we have also assumed Lorentz invariance
of the vacuum: 
$\delta\dot{\phi}_{\ell}\simeq\delta\partial_{i}\phi_{\ell}\simeq\delta\phi_{\ell}/{\ell}$,
and used the Wick rotation to evaluate the expectation value in (\ref{eq:assumption}),
so that $X_{\ell}<0$. This is the second of our assumptions. This
assumption can also be motivated by dimensional regularization. Indeed,
$\left\langle 0\right|X\left|0\right\rangle \neq0$ but to calculate
this quantity one has to use some regularization which is compatible
with the Lorentz symmetry of $\left|0\right\rangle $. Usually the
Wick rotation works for interacting theories with energies bounded
from below. Moreover, light-like $X_{\ell}=0$, would not change
the fluctuations at all, whereas timelike $X_{\ell}=\left(\delta\phi/\ell\right)^{2}$
intuitively implies some flow or evolution in time which should not
be the case in a static situation. 

Further, we assume that that the vacuum $\left|0\right\rangle $ is
standard so that $\left\langle 0\right|\hat{\phi}_{\ell}\left|0\right\rangle =\left\langle 0\right|\hat{p}_{\ell}\left|0\right\rangle =0$.
In vacuum $\left|0\right\rangle $ the product of uncertainties is
smaller than in any other energy eigenstate $\left|E\right\rangle $.
We can express the product of fluctuations as 

\begin{equation}
\delta\phi\left(\ell\right)\cdot\delta p\left(\ell\right)=\frac{\hbar\mathscr{D}_{0}}{2}\cdot\Omega\left(\ell\right)\cdot\ell^{-d}\geq\frac{\hbar\mathscr{D}_{0}}{2}\cdot\ell^{-d}\,,\label{eq:Fluctuations_In_Vacuum}
\end{equation}

where $\Omega\left(\ell\right)$ is a function, $\Omega\left(\ell\right)\geq1$,
fixed by properties of the vacuum state which in turn are fixed by
the theory. If $\Omega\left(\ell\right)=1$ the vacuum saturates the
Heisenberg uncertainty relation by absolute minimization of the product
of fluctuations. Now it is easy to see that $\mbox{d}\Omega/\mbox{d}\ell$
should be negative in UV or asymptotically approach zero. Indeed,
if $\Omega\left(\ell\right)$ were increasing, $\mbox{d}\Omega/\mbox{d}\ell>0$, it would imply that
in deep UV the system would violate the uncertainty relation $\Omega\left(\ell\right)\geq1$.
Otherwise, one has to assume that there is another scale different
from $M_{\text{s}}^{-1}$ where $\mbox{d}\Omega/\mbox{d}\ell$ changes
the sign. If $\mbox{d}\Omega/\mbox{d}\ell<0$ in UV it can change
the sign and approach zero in IR for $\ell\gtrsim M_{\text{s}}^{-1}$. 

Further, by plugging the characteristic momentum into the Heisenberg
uncertainty relation (\ref{eq:Fluctuations_In_Vacuum}) we obtain
\begin{equation}
\mathscr{L}_{X}\left(X_{\ell}\right)\delta\phi^{2}\left(\ell\right)\simeq\frac{\hbar\mathscr{D}_{0}}{2}\cdot\Omega\left(\ell\right)\cdot\ell^{1-d}\,.\label{FirstIntegral}
\end{equation}
This yields an implicit function $\delta\phi\left(\ell\right)$
for general theories with Lagrangians $\mathscr{L}\left(X\right)$
for which our assumptions (\emph{classicalization}) are fulfilled.
Below in this letter, we find $\delta\phi\left(\ell\right)$
analytically for DBI-like theories, assuming that $\Omega\left(\ell\right)$
is a slowly varying function of the scale $\ell$, so that $\mbox{d}\ln\Omega/\mbox{d}\ln\ell\ll1$.
In most cases it is not possible to solve this algebraic equation
(\ref{FirstIntegral}) exactly even for a given $\Omega\left(\ell\right)$.
Nevertheless, we can obtain important information e.g. the slope of
the fluctuations by differentiating this equation (\ref{FirstIntegral})
with respect to the length scale $\ell$ to obtain
\begin{equation}
\frac{\mbox{d}\delta\phi}{\mbox{d}\ell}=\frac{\delta\phi}{\ell}\cdot\left[\frac{v^{2}-d+\mbox{d}\ln\Omega/\mbox{d}\ln\ell}{v^{2}+1}\right]\,,\label{SlopeOfFluctuations}
\end{equation}
where we have introduced the notation 
\begin{equation}
v^{2}=v^{2}\left(X_{\ell}\right)=1+\left.\frac{2X\mathscr{L}_{XX}}{\mathscr{L}_{X}}\right|_{X=X_{\ell}}\,.\label{Speed}
\end{equation}
It is important to note that any dependence on the window-function
parameter $\mathscr{D}_{0}$ disappeared from Eq. (\ref{SlopeOfFluctuations}),
which can be also considered as a first-order differential equation
for $\delta\phi\left(\ell\right)$. In that case, $\mathscr{D}_{0}$
is the integration constant. The physical meaning of the function
$v(X)$ is the (radial) speed of propagation for small perturbations
around a \textit{classical static background} $\phi\left(\mathbf{x}\right)$
at a point where $X=-\frac{1}{2}\partial_{i}\phi\partial_{i}\phi$
coincides with the characteristic $X_{\ell}$ given by (\ref{X})
for quantum fluctuations. Indeed, it is well known (see Ref.\,\cite{ArmendarizPicon:2005nz})
that for the spacelike classical backgrounds the (radial) sound speed
is given by the formula\textit{ }(\ref{Speed}) which is the\textit{
inverse} of the one derived in Ref.\,\cite{Garriga:1999vw} for cosmological
backgrounds in the context of general k-\textit{essence} theories
\cite{ArmendarizPicon:1999rj,ArmendarizPicon:2000dh,ArmendarizPicon:2000ah}. 

It is worth mentioning that for $\mbox{d}\ln\Omega/\mbox{d}\ln\ell=0$
the differential equation (\ref{SlopeOfFluctuations}) is invariant
under simultaneous rescaling $\delta\phi\rightarrow s\delta\phi$
and $\ell\rightarrow s\ell$, while for a constant $v$ the rescaling
parameters are independent. However, any solution of this equation
given by (\ref{FirstIntegral}) with a \textit{fixed} $\mathscr{D}_{0}$
is not invariant under this rescaling. 

It is also convenient to write equation (\ref{SlopeOfFluctuations})
as 
\begin{equation}
\frac{\mbox{d}\delta\phi}{\mbox{d}\ell}=\frac{\delta\phi}{\ell}\left[\frac{1-d}{2}+\frac{\left(1+d\right)\left(v^{2}-1\right)+2\ell\mbox{d}\Omega/\Omega\mbox{d}\ell}{2\left(v^{2}+1\right)}\right]\,.
\end{equation}
Thus quantum fluctuations can be suppressed with respect to the weekly
coupled case, provided either 
\begin{equation}
v^{2}>1-\frac{2}{1+d}\cdot\frac{\mbox{d}\ln\Omega}{\mbox{d}\ln\ell}\geq1\,,
\end{equation}
or $v^{2}<-1$. While for an absolute suppression, with \textit{red}
tilt, $\mbox{d}\delta\phi/\mbox{d}\ell>0$, one has to require either
a stronger condition 
\begin{equation}
v^{2}>d-\frac{\mbox{d}\ln\Omega}{\mbox{d}\ln\ell}\geq d\,,\label{SuperluminalCondition}
\end{equation}
or again that $v^{2}<-1$. 

The first case, $v^{2}>1$, implies that the theory possesses classical
static backgrounds $\phi\left(\mathbf{x}\right)$ around which small
perturbations propagate faster than light. Thus a \textit{quasiclassical
attractive} interaction is only possible for Nambu-Goldstone bosons
allowing for superluminality. The second case, $v^{2}<-1$, implies
that there are backgrounds $\phi\left(\mathbf{x}\right)$ which are
absolutely unstable - they possess gradient instabilities. Faster-than-light
propagation around nontrivial backgrounds $\phi\left(\mathbf{x}\right)$
does not imply an immediate inconsistency, see \cite{ArmendarizPicon:2005nz,Babichev:2007dw,Bruneton:2006gf,Kang:2007vs,Geroch:2010da},
but rather signals the absence of the usual Lorentz invariant and
local Wilsonian UV-completion by other fields appearing in UV\cite{Adams:2006sv}.
However, exactly such theories may be able to \textit{classicalize},
as it was argued in \cite{Dvali:2012zc,Dvali:2011nj}. Another well
known problem associated with such theories is that the relation between
$p$ and $\dot{\phi}$ can be multivalued \cite{Aharonov:1969vu},
for a recent interesting proposal of how to deal with this problem
see \cite{Shapere:2012nq,Shapere:2012yf}. 

Do the fluctuations calculated above correspond to the minimal energy
i.e. vacuum? Similarly to the well known estimation for the lowest
energy level of Hydrogen we can estimate the size of fluctuations
minimising the Hamiltonian density $\mathcal{H}\left(p,\left(\nabla\phi\right)^{2}\right)=p\dot{\phi}-\mathscr{L}$
where the characteristic momentum is $p=\delta p$ and given by the
Heisenberg uncertainty relation (\ref{HeisenbergUR-1}). Further we
will assume that the level of self-interaction for each mode is much
stronger than the coupling to other modes. The condition for the extremum
is 
\begin{equation}
\frac{\partial\mathcal{H}}{\partial\delta\phi}=\dot{\phi}\frac{\partial p}{\partial\delta\phi}+\frac{1}{2}\mathscr{L}_{X}\frac{\partial\left(\nabla\phi\right)^{2}}{\partial\delta\phi}\simeq0\,.
\end{equation}
One can check that, under our assumptions, (\ref{FirstIntegral})
with some $\Omega$ is a solution of this equation. 
\section{Lagrange multiplier renormalises $\hbar$ }

It is more intuitive to deal with systems quadratic in derivatives.
We reformulate the theory using a Lagrange multiplier field $\lambda$
and the Legendre transformation in the following way 
\begin{equation}
S=\int\mbox{d}^{1+d}x\,\left(\lambda\cdot X-V\left(\lambda\right)\right)\,.
\end{equation}
The equations of motion for $\lambda$ would give us 
$X=V_{\lambda},$
providing an implicit function $\lambda\left(X\right)$. Therefore
the ``potential'', $V\left(\lambda\right)$, describing the original
theory (\ref{ActionGeneral}) is given by the relation 
\begin{equation}
V\left(\lambda\left(X\right)\right)=\lambda\left(X\right)\cdot X-\mathscr{L}\left(X\right)\,.\label{PotentialV}
\end{equation}
Differentiating the relation (\ref{PotentialV}) with respect to $X$
one obtains $\lambda=\mathscr{L}_{X}$.
If we assume a negligibly weak correlation between $\lambda$ and
$\dot{\phi}$, then 
\begin{equation}
\delta p^{2}\left(\ell\right)=\bigl\langle(\lambda\dot{\phi})^{2}\bigr\rangle_{\ell}\simeq\bigl\langle\lambda^{2}\bigr\rangle_{\ell}\bigl\langle\dot{\phi}^{2}\bigr\rangle_{\ell}=\bigl\langle\lambda^{2}\bigr\rangle_{\ell}\cdot\delta\dot{\phi}^{2}\left(\ell\right)\,.
\end{equation}
Note that $\bigl\langle\lambda\bigr\rangle_{\ell}\neq0$. If we assume
weak coupling, or the \textit{quasiclassical} approximation i.e. negligible
dispersion of $\lambda$, then $\bigl\langle\lambda^{2}\bigr\rangle_{\ell}\simeq\bigl\langle\lambda\bigr\rangle_{\ell}^{2}$.
Further we use the notation $\bigl\langle\lambda\bigr\rangle_{\ell}\equiv\lambda\left(\ell\right)$
omitting sometimes the dependence on $\ell$, and writing just $\lambda$.
The standard Heisenberg uncertainty relation (\ref{HeisenbergUR-1})
takes the form 
\begin{equation}
\delta\phi\left(\ell\right)\cdot\delta\dot{\phi}\left(\ell\right)\geq\frac{\mathscr{D}_{0}}{2}\left(\frac{\hbar}{\lambda\left(\ell\right)}\right)\cdot\ell^{-d}\,.
\end{equation}
If we assume that the Lorentz-invariant vacuum saturates the inequality and drop the unimportant numerical factor $\mathscr{D}_{0}/2$
the quantum fluctuations are given by 
\begin{equation}
\delta\phi\left(\ell\right)\simeq\left(\frac{\hbar}{\lambda\left(\ell\right)}\right)^{1/2}\cdot\ell^{\left(1-d\right)/2}\,.\label{QF}
\end{equation}
Thus quantum fluctuations $\delta\phi\left(\ell\right)$ are suppressed
with respect to the standard result, provided $\lambda\left(\ell\right)=\bigl\langle\mathscr{L}_{X}\bigr\rangle_{\ell}>1$
parametrically. If, moreover, $\lambda\left(\ell\right)$ evolves
with $\ell$ so that, the shorter is $\ell$ the larger is $\lambda\left(\ell\right)$,
then the shorter is $\ell$ the more suppressed are quantum fluctuations.
Thus \textit{classicalization} occurs when $\mbox{d}\lambda/\mbox{d}\ell<0$.

Further, just from dimensional reasons and again assuming weak coupling,
or the \textit{quasiclassical} approximation, we obtain %
\footnote{Here we could also estimate that $V\left(\lambda\left(\ell\right)\right)\simeq\left\langle V\right\rangle _{\ell}\simeq\ell^{-\left(1+d\right)}$.
However, this estimation leads to the dependence of the final result
not only on the derivatives of the Lagrangian $\mathscr{L}$, but
also on its value and in particular on the possible cosmological constant.
Clearly vacuum fluctuations of the scalar field should not depend
on the value of the cosmological constant provided we ignore the effects
related to spacetime curvature.%
}
\begin{equation}
\lambda\left(\ell\right)V_{\lambda}\left(\lambda\left(\ell\right)\right)\simeq\bigl\langle\lambda V_{\lambda}\bigr\rangle_{\ell}\simeq\ell^{-\left(1+d\right)}\,.
\end{equation}
Differentiating this expression with respect to $\ell$ yields 
\begin{equation}
\frac{\mbox{d}\lambda}{\mbox{d}\ell}\left(V_{\lambda}+\lambda V_{\lambda\lambda}\right)=-\left(1+d\right)\frac{\lambda V_{\lambda}}{\ell}\,,
\end{equation}
so that, noting that $V_{\lambda\lambda}=1/\mathscr{L}_{XX}$, we
can recast this equation in the form 
\begin{equation}
\frac{\mbox{d}\lambda}{\mbox{d}\ell}=-\left(1+d\right)\cdot\frac{\lambda}{\ell}\cdot\left(\frac{v^{2}-1}{v^{2}+1}\right)\,,\label{dlambdaOverdl}
\end{equation}
where the velocity $v$ is given by the formula (\ref{Speed}). From
this equation we infer that to have larger $\lambda$ for smaller
scales it is necessary that either $v^{2}<-1$ or $v^{2}>1$. But
it is exactly this behavior of $\lambda\left(\ell\right)$ which is
required to suppress $\delta\phi\left(\ell\right)$ relatively to
the standard case, see Eq. (\ref{QF}) . Thus we confirm in that way
our previous analysis. Finally differentiating (\ref{QF}) with respect
to $\ell$ and using (\ref{dlambdaOverdl}) we reproduce the result
(\ref{SlopeOfFluctuations}) for slowly-varying $\Omega$. 

\section{($a$)DBI \label{sec:DBI-and-anti-DBI}}

Here, to illustrate our analysis, we consider theories described by
the Lagrangian 
\begin{equation}
\mathscr{L}=\sigma M_{\text{s}}^{d+1}\left[\sqrt{1+\frac{2X}{\sigma M_{\text{s}}^{d+1}}}-1\right]\,,\label{action}
\end{equation}
where $M_{\text{s}}$ is a strong-coupling scale, and $\sigma=\pm1$.
The case $\sigma=-1$, corresponds to the standard Dirac-Born-Infeld
(DBI) theory, where small perturbations always propagate with the
speed less or equal to the speed of light - i.e. they are never\textit{
superluminal}. Whereas the case $\sigma=+1$ corresponds to the so-called
anti-DBI (aDBI) theory which was introduced in \cite{Mukhanov:2005bu,Babichev:2006vx,Babichev:2007wg}
and for which small perturbations are never \textit{subluminal}. aDBI
was extensively studied as an example of a theory which can classicalize,
see \cite{Dvali:2010jz,Rizos:2011wj,Dvali:2012zc,Rizos:2012qs,Alberte:2012is}.
For small derivatives the system approaches a free massless scalar
field. Another useful property of (a)DBI is that the real equations
of motion are always hyperbolic (i.e. there is no gradient instability).
Moreover, (a)DBI satisfy Null Energy Condition -- they do not have ghosts
and possess nonnegative energy density, see e.g. \cite{Rizos:2012qs}.
The unphysical region of phase space corresponding to imaginary observables
is separated by a barrier of infinite positive energy density. \\

For this theory the saturated uncertainty relation (\ref{FirstIntegral})
takes the form 
\begin{equation}
\frac{\left(\ell^{\left(d-1\right)/2}\,\delta\phi\right)^{2}}{\sqrt{1-2\sigma\left(\ell^{\left(d-1\right)/2}\,\delta\phi\right)^{2}/\left(\ell M_{\text{s}}\right)^{d+1}}}\simeq1\,,
\end{equation}
 where we have omitted the irrelevant window-function factor $\mathscr{D}_{0}/2$
which would also only change the normalisation of $\delta\phi\left(\ell\right)$
and $M_{\text{s}}$ by a factor $\mathcal{O}\left(1\right)$ but not
the scaling with $\ell$. The above equation has the following solution
\begin{equation}
\delta\phi^{2}\left(\ell\right)=\frac{\ell^{1-d}}{\left(\ell M_{\text{s}}\right)^{d+1}}\left[\sqrt{1+\left(M_{\text{s}}\ell\right)^{2\left(d+1\right)}}-\sigma\right]\,.\label{DBIFluct}
\end{equation}
In IR, $\ell\gg M_{\text{s}}^{-1}$, one obtains 
\begin{equation}
\delta\phi\left(\ell\right)\simeq\ell^{\left(1-d\right)/2}\cdot\left(1-\frac{\sigma}{2}\left(\ell M_{\text{s}}\right)^{-\left(d+1\right)}\right)\,.
\end{equation}
The leading term reproduces the standard result. However, the next-to-leading 
term reveals that for aDBI the fluctuations are slightly suppressed
whereas for the DBI the fluctuations are slightly enhanced. 

In the UV, $\ell\ll M_{\text{s}}^{-1}$, quantum fluctuations already
in the leading order depend on the sign $\sigma$. Namely, for superluminal
aDBI, ($\sigma=+1$) the quantum fluctuations have a \textit{red}
spectrum: 
\begin{equation}
\delta\phi_{\text{aDBI}}\left(\ell\right)\sim M_{\text{s}}^{\left(d+1\right)/2}\cdot\ell\,,\label{fluct_aDBI}
\end{equation}
whereas for the standard DBI ($\sigma=-1$) the spectrum is \textit{blue}:
\begin{equation}
\delta\phi_{\text{DBI}}\left(\ell\right)\sim M_{\text{s}}^{-\left(d+1\right)/2}\cdot\ell^{-d}\,.
\end{equation}
Thus, for scales shorter than the strong-coupling scale $M_{\text{s}}^{-1}$,
quantum fluctuations in aDBI become softer - the behavior which one
would expect for those systems which \textit{classicalize}. Here it
should be mentioned that a \textit{classicalon} of scale $\ell$ would
have the maximal field value given by (\ref{fluct_aDBI}). 

For the corresponding conjugate momenta in the UV, the Heisenberg
uncertainty relation (\ref{HeisenbergUR-1}) yields
\begin{equation}
\delta p_{\text{aDBI}}\left(\ell\right)\sim M_{\text{s}}^{-\left(d+1\right)/2}\cdot\ell^{-\left(d+1\right)}\,, \ \  \delta p_{\text{DBI}}\left(\ell\right)\sim M_{\text{s}}^{\left(d+1\right)/2}\,.
\end{equation}
Thus, there is a duality between these two theories in the sense of 
\begin{equation}
\delta p_{\text{aDBI}}\left(\ell\right)\sim\frac{\delta\phi_{\text{DBI}}\left(\ell\right)}{\ell}\ , \  \ \delta p_{\text{DBI}}\left(\ell\right)\sim\frac{\delta\phi_{\text{aDBI}}\left(\ell\right)}{\ell}\,.
\end{equation}
Moreover, the vacuum state is squeezed in orthogonal directions for
these theories: 
\begin{equation}
 \frac{\delta p_{\text{aDBI}}}{\delta\phi_{\text{aDBI}}/\ell}\sim\left(\ell M_{\text{s}}\right)^{-\left(d+1\right)} \sim \frac{{\delta\phi_{\text{DBI}}/\ell}}{\delta p_{\text{DBI}}}\gg1\ .
\end{equation}
It is also interesting to note that one does not obtain these estimations
for the spectrum $\delta\phi\left(\ell\right)$ by assuming the Euclidian
action to be of order unity, i.e. $\hbar$. Presumably this happens,
because classically the canonical momentum is restricted for the aDBI,
see e.g. \cite{Rizos:2011wj,Rizos:2012qs}. 

To conclude this section a cautionary remark is necessary. The usual
DBI can be UV-completed in a standard Wilsonian way. Therefore the
validity of the quasiclassical approximation is rather questionable
in this case and we put much less value on the estimations obtained
for the usual DBI in this way. 
\section{Conclusions}

We have demonstrated that the quasiclassical suppression of quantum
fluctuations of Nambu-Goldstone bosons is only possible for theories
which either possess catastrophically unstable backgrounds or allow
for superluminality. This finding supports the analysis performed
before in \cite{Dvali:2011nj,Dvali:2012zc}. 

It would be very interesting to study whether these results can be
generalised to other systems and quantum states. 
\begin{acknowledgments}
It is great pleasure to thank A. Barvinsky, M. Buican, G. D'Amico,
S. Dubovsky, G. Dvali, A. Frolov, V. Frolov, C. Gomez, S. Hofmann, N. Kaloper, A. Khmelnitsky,
D. Malyshev, V. Mukhanov, O. Pujol\`as, O. Ruchayskiy, V. Rubakov,
M. Sasaki, I. Sawicki, S. Sibiryakov, A. Starobinsky, N. Tetradis,
V. Vanchurin, A. Vilenkin and S. Winitzki for very useful valuable
discussions and criticisms. That does not mean that they necessarily
agree with anything written in this work. 
The first version of this paper  was prepared during my visits to to the
ASC, Ludwig-Maximilians-Universit\"at M\"unchen and SITP, Stanford
University, while the second version was written during the YITP-T-12-3 workshop at the YITP, Kyoto University for whose warm hospitality and atmosphere 
I am very thankful.  
\end{acknowledgments}
\bibliographystyle{utphys}
\phantomsection\addcontentsline{toc}{section}{\refname}\bibliography{Qfluct}

\end{document}